# Study of the effect of quantum noise on the accuracy of the Schrödinger equation simulation on a quantum computer using the Zalka-Wiesner method

Yu.I. Bogdanov[*], N.A. Bogdanova, D.V. Fastovets, V.F. Lukichev

Valiev Institute of Physics and Technology of Russian Academy of Sciences, Russia, Moscow

## ABSTRACT

The study of the effect of quantum noise on the accuracy of modeling quantum systems on a quantum computer using the Zalka-Wiesner method is carried out. The efficiency of the developed methods and algorithms is demonstrated by the example of solving the nonstationary Schrödinger equation with allowance for quantum noise for a particle moving in the Pöschl–Teller potential. The analysis and prediction of accuracy for the Zalka-Wiesner method are carried out taking into account the complexity of the quantum system and the strength of quantum noise. In our opinion, the considered problem provides a useful test for assessing the quality and efficiency of quantum computing devices on various physical platforms currently being developed. The obtained results are essential for the development of high-precision methods for controlling the technologies of quantum computations.

## 1. INTRODUCTION

Quantum computers allow us to achieve exponential acceleration during solving certain practical problems. One of the most important such problems is the problem of modeling quantum systems. The need for high-precision modeling of the quantum systems evolution of various nature arises in such scientific fields as physics [1], quantum chemistry [2] and computer technologies [3]. The most direct way to solve problems in the case of quantum mechanical systems is to find a solution to the nonstationary Schrödinger equation. This equation, of course, can be solved numerically using a classical computer. However, in this case, to carry out high-precision modeling of real practically important problems, an exponentially large amount of resources (in terms of memory and execution time) will be required. On the other hand, a quantum computer can be used to simulate the evolution of a quantum system. It is shown that a quantum computer can effectively solve the nonstationary Schrödinger equation using the Zalka-Wiesner algorithm [4,5], which is one of the main methods for analyzing quantum systems using quantum computing.

The considered algorithm makes it possible to use the resources of quantum computing to simulate the evolution of a quantum system described by an arbitrary Hamiltonian. The Zalka-Wiesner algorithm is actively studied by various scientific groups. The efficiency of this algorithm on a small number of qubits has been proven and demonstrated numerically [6], and it has also been tested using an IBM cloud quantum processor [7]. The universality of this algorithm allows us to apply it to various quantum systems, including many-body systems with Coulomb interaction [8] and systems of quantum harmonic oscillators [9].

Note that several modifications of this algorithm have been developed. For example, the adaptation of the Zalka-Wiesner algorithm to the simulation of chemical reactions is described [10]. The authors of this work pay special attention to the preparation of the required quantum state, which is an important problem in the implementation of some quantum algorithms. In this work, approbation is carried out during simulating a small number of atoms (4-5), and methods for calculating chemically significant observables are also presented. Another work [11] is devoted to a modification of the Zalka-Wiesner algorithm, which exceeds the explicit difference scheme in accuracy, but requires exponential memory costs. The developed algorithm makes it possible to simulate the dynamics in real time, which was confirmed in numerical experiments.

---

[*] bogdanov_yurii@inbox.ru

The main obstacle to the development of quantum computers and quantum simulators is associated with the influence of decoherence and quantum noise on the dynamics of multi-qubit quantum systems [12]. The efficiency of the Zalka-Wiesner algorithm taking into account the effect of decoherence is an urgent and not well-studied problem. Such studies, in our opinion, are critically important for the practical implementation of the considered quantum algorithm. The obtained results can be used to improve the quality of the algorithm both at the software and hardware levels. The study of the Zalka-Wiesner algorithm is the subject of this work. Section 2 describes the general idea of the Zalka-Wiesner method. It is important to note that each step in the evolution of the quantum state of the qubit register is associated with the usage of the direct and inverse Fourier transforms. Thus, the accuracy of the Fourier transform in the quantum register limits the final accuracy of solving the nonstationary Schrödinger equation. An estimate of the quantum noise effect on the quantum Fourier transform using an approach based on the theory of quantum operations is considered in Section 3. Section 4 presents numerical modeling examples of the approximative quantum Fourier transform. Section 5 presents the approbation of the developed methods and algorithms using the example of considering the dynamics of a quantum particle in the Pöschl-Teller potential. Section 6 presents conclusions of the work.

## 2. GENERAL IDEA OF THE ZALKA-WIESNER METHOD

For simplicity, let consider the one-dimensional dynamics of a quantum particle. Formal multidimensional many-particle system generalization is carried out directly. The total evolution time $t$ is divided into a sufficient number of intervals $n_t$. Simulation is carried out on a separate, small interval $\Delta t = t/n_t$ and repeated $n_t$ times.

The Zalka-Wiesner algorithm begins with taking into account the influence of the potential energy operator in the coordinate representation. Then, momentum representation transition is made using Fourier transform, where the kinetic energy operator is taken into account. Then, we return to the original coordinate representation using inverse Fourier transform. Thus, the unitary evolution operator in coordinate space on a small time interval is

$$U_{\Delta t} = IQFT \cdot \exp\left(-\frac{i}{\hbar}\frac{p^2}{2m}\Delta t\right) QFT \cdot \exp\left(-\frac{i}{\hbar}V(x)\Delta t\right). \tag{1}$$

Here $\exp\left(-\frac{i}{\hbar}V(x)\Delta t\right)$ - evolution operator in coordinate space, defined by potential function $V(x)$; $QFT$ (Quantum Fourier Transform) - a quantum Fourier transform that provides a transition from a coordinate representation to momentum representation; $\exp\left(-\frac{i}{\hbar}\frac{p^2}{2m}\Delta t\right)$ - evolution operator in momentum space, defined by the kinetic energy operator $\frac{p^2}{2m}$; $IQFT$ (Inverse Quantum Fourier Transform) – an inverse quantum Fourier transform that provides a transition from a momentum representation to coordinate representation.

The direct quantum Fourier transform ($QFT$) of the wave function, which ensures the transition from the coordinate representation to the momentum one, is

$$\tilde{\psi}(p) = \frac{1}{\sqrt{2\pi\hbar}} \int \exp\left(-\frac{i}{\hbar}px\right)\psi(x)dx.$$

The inverse quantum Fourier transform ($IQFT$), which provides the transition from the momentum representation to the coordinate one, is

$$\psi(x) = \frac{1}{\sqrt{2\pi\hbar}} \int \exp\left(\frac{i}{\hbar}px\right)\tilde{\psi}(p)dp.$$

The presence of a quantum register of $n$ qubits provides $N = 2^n$ sampling points and corresponds to the evolution of the corresponding discretized quantum system in the Hilbert space with dimension $N$.

The evolution representation form (1) is a good approximation for sufficiently large $n_t$ value because of the Trotter formula

$$\lim_{n_t \to \infty} \left[ \exp(-iAt/n_t) \exp(-iBt/n_t) \right]^{n_t} = \exp\left[-i(A+B)t\right]. \quad (2)$$

It is known that if the operators do not commute, then

$$\exp\left[-i(A+B)t\right] \neq \exp(-iAt)\exp(-iBt),$$

at finite times $t$. However, this kind of relation is fulfilled for small times $\Delta t$ because of Trotter's formula (2). In our case, the role $A$ and $B$ are played by the operators of kinetic and potential energy, respectively. In this work, we used two forms of the Trotter formula [12], which differ in the accuracy order:

$$\exp\left[-i(A+B)\Delta t\right] = \exp[-iA\Delta t]\exp[-iB\Delta t] + O\left((\Delta t)^2\right),$$

$$\exp\left[-i(A+B)\Delta t\right] = \exp[-iA\Delta t/2]\exp[-iB\Delta t]\exp[-iA\Delta t/2] + O\left((\Delta t)^3\right).$$

The first of these formulas will be called default, and the second (more precise) - modified.

## 3. QUANTUM NOISE INFLUENCE ON THE QUANTUM FOURIER TRANSFORM: AN APPROACH BASED ON QUANTUM OPERATION THEORY

An ideal quantum circuit is a sequential set of unitary quantum transforms $\{U_i\}$. Operations $U_i$ usually act on one, two, or, less commonly, three qubits. For real physical systems, the implementation of each transform $U_i$ implies the presence of the corresponding Hamiltonian performing the operation $U_i = e^{-iH_i t_i / \hbar}$ in time $t_i$. Thus, for further consideration, it suffices to use the operators $U_i$ specification. The problem of modeling the quantum circuit operation is to find the final state $U_s \cdot U_{s-1} \cdot \ldots \cdot U_2 \cdot U_1 | \psi_0 \rangle$ of the system from the initial state $| \psi_0 \rangle$.

To describe open quantum systems affected by quantum noise, it is necessary to use the density matrix formalism. In this case, quantum states transforms expand from unitary gates to so-called quantum operations [13, 14]. The action of an arbitrary quantum operation on a quantum state in the form of a density matrix can be represented in operator sum form: $\varepsilon(\rho) = \sum_i E_i \rho E_i^\dagger$, where the operators $E_i$ satisfy the normalization condition (norm preservation) and are called Kraus operators.

In terms of state vectors, noise can be introduced using the Monte Carlo method. For example, an arbitrary one-qubit gate $U_{ideal}$ can be replaced with a noisy analog as follows

$$U = U_{ideal} \cdot V_{noise}, \quad V_{noise} = \begin{pmatrix} \cos(e\xi) & \sin(e\xi) \\ -\sin(e\xi) & \cos(e\xi) \end{pmatrix}, \quad (3)$$

where $\xi$ - normally distributed random variable ($\xi \sim N(0,1)$) and $e$ – the error level. This representation of the noisy gate turns out to be very useful and will be used by us further. The controlled phase rotation noisy operator used in the quantum Fourier transform can be represented in a similar way

$$U_\theta = U_\theta^{ideal} \cdot V_\theta^{noise} = \begin{pmatrix} 1 & 0 & 0 & 0 \\ 0 & 1 & 0 & 0 \\ 0 & 0 & 1 & 0 \\ 0 & 0 & 0 & e^{-i(\theta + e \cdot \xi)} \end{pmatrix}, \quad (4)$$

where $U_\theta^{ideal}$ - the ideal phase rotation operator and $V_\theta^{noise}$ – corresponding noise operator. For simplicity, we will characterize the noise level in formulas (3) and (4) by the same parameter $e$.

It can be shown [15] that the noisy transform $V_{noise}$ for single-qubit operators (3) is statistically equivalent to the following set of Kraus operators

$$E_1 = \sqrt{\lambda_1} \begin{pmatrix} 1 & 0 \\ 0 & 1 \end{pmatrix}, \quad E_2 = \sqrt{\lambda_2} \begin{pmatrix} 0 & 1 \\ -1 & 0 \end{pmatrix},$$

where

$$\lambda_1 = \frac{1}{2}\left(1 + \exp(-2e^2)\right), \quad \lambda_2 = \frac{1}{2}\left(1 - \exp(-2e^2)\right). \quad (5)$$

Similarly, for a noisy controlled phase shift operator $V_\theta^{noise}$ from (4) we have

$$E_1 = \begin{pmatrix} 1 & 0 & 0 & 0 \\ 0 & 1 & 0 & 0 \\ 0 & 0 & 1 & 0 \\ 0 & 0 & 0 & \sqrt{P} \end{pmatrix}, \quad E_2 = \begin{pmatrix} 0 & 0 & 0 & 0 \\ 0 & 0 & 0 & 0 \\ 0 & 0 & 0 & 0 \\ 0 & 0 & 0 & \sqrt{1-P} \end{pmatrix}, \quad (6)$$

where $P = \exp(-e^2)$.

Let the initial state of the system arriving at the input of the simulated circuit be pure. Then the corresponding density matrix has unit rank. Each Kraus operator specifies one of the quantum system evolution alternative paths. Therefore, each quantum operation increases the rank of the density matrix exponentially up to the maximum possible rank $r = 2^n$, where $n$ is the number of qubits in the system. To reduce the number of required computing resources, at each step, we will approximate the resulting density matrix by a density matrix of a lower rank (keeping only a fixed number of maximum eigenvalues and zeroing the rest). Such approximation preserves the positive-definiteness of the density matrix. However, density matrix trace decreases, which corresponds to a decrease of states evolving along the "correct" path.

Direct application of the unit rank approximation to the quantum Fourier transform leads to the following estimate, which specifies the transform accuracy for a random state [15]:

$$F_{QFT} = P_H^n P_R^{n(n-1)/8}, \quad (7)$$

where $P_H = \frac{1}{2}\left(1 + \exp(-2e^2)\right)$, $P_R = \exp(-e^2)$ - the no-error probabilities of the Hadamard transform and the conditional phase shift, respectively.

The presented based on the fact that the Fourier transform contains $n$ Hadamard gates and $n(n-1)/2$ two-qubit phase transforms. In this case, for a random input state, uniformly distributed according to the Haar measure, the two-qubit phase transform affects only one fourth of the total number of amplitudes of the random state at each step. Therefore, the two-qubit noise generates an exponent of power $n(n-1)/8$. This formula provides a fairly rough fidelity estimation.

The considered estimate can be significantly improved by transforming to the principal axes of the conditional phase shift Kraus operators (6). It is well known that Kraus operators are defined up to a wide unitary arbitrariness. The diagonals of the matrices become orthogonal with respect to each other by choosing a suitable unitary transform. The new representation of Kraus operators has the form [15]:

$$\tilde{E}_1 = \frac{1}{\sqrt{1+f^2}} \begin{pmatrix} 1 & 0 & 0 & 0 \\ 0 & 1 & 0 & 0 \\ 0 & 0 & 1 & 0 \\ 0 & 0 & 0 & \sqrt{P}+f\sqrt{1-P} \end{pmatrix}, \quad \tilde{E}_2 = \frac{1}{\sqrt{1+f^2}} \begin{pmatrix} -f & 0 & 0 & 0 \\ 0 & -f & 0 & 0 \\ 0 & 0 & -f & 0 \\ 0 & 0 & 0 & -f\sqrt{P}+\sqrt{1-P} \end{pmatrix}. \quad (8)$$

where $P = \exp(-e^2)$, $f = \dfrac{\sqrt{1+3P}-P-1}{\sqrt{P(1-P)}}$.

Kraus operators (6) and (8) are unitarily equivalent and, therefore, define the same quantum operation. Then, an improved accuracy estimate of the quantum Fourier transform of a random state is

$$F_{QFT} = P_H^n \tilde{P}_R^{n(n-1)/8}. \quad (9)$$

where $\tilde{P}_R = \dfrac{\left(\sqrt{P}+f\sqrt{1-P}\right)^2}{\left(1+f^2\right)^4}$.

The developed approximate approach makes it possible to predict the effect of quantum noise on the multi-qubit quantum circuits accuracy, which have not yet been implemented "in hardware" and which cannot be simulated on any modern or promising classical computers.

## 4. NUMERICAL SIMULATION OF THE APPROXIMATE QUANTUM FOURIER TRANSFORM

The angle $\theta$ in (4) is set by the conditional phase rotation operator $CR_k$ and is equal to $\theta = \dfrac{2\pi}{2^k}$ ($2 \leq k \leq n$). If $k$ is large, then the angle $\theta$ becomes such a small value that cannot be realized in any modern or promising experiments. If we implement $QFT$ only for $k \leq k_0$, where is the $k_0$ depth of approximation, we will obtain the Approximate Quantum Fourier Transform ($AQFT$). The optimal $k_0$, determined by the errors level, can be estimated by the formula $k_0^{optim} = \log_2\left(\dfrac{2\pi}{e}\right)$. This optimum condition corresponds to the situation with equal signal $\theta = \dfrac{2\pi}{2^{k_0}}$ and noise $e$. The $k_0^{optim}$ parameter sets the lower limit for the use of the corresponding gate $CR_k$ since the signal prevails over the noise. Noise prevails above this limit, so it is unprofitable to use the corresponding gate.

Figure 1 shows the results of numerical simulations by the Monte Carlo method for the dependence of the fidelity loss from the AQFT depth (1000 uniformly distributed states according to the Haar measure were considered at each point). Left figure corresponds to the noise level $e = 0.05$ and $n = 12$ qubits. In this case $k_0^{optim} \approx 6.973$, which is in good agreement with the numerical experiment (the fidelity loss is significantly lower when we make phase rotations "according to the reduced program" at $k_0$=7, in comparison with the case of the full QFT, when $k_0$=12. Similarly, on the right figure, we see that the fidelity loss has a minimum at $k_0$=9, in this case $k_0^{optim} \approx 9.295$.

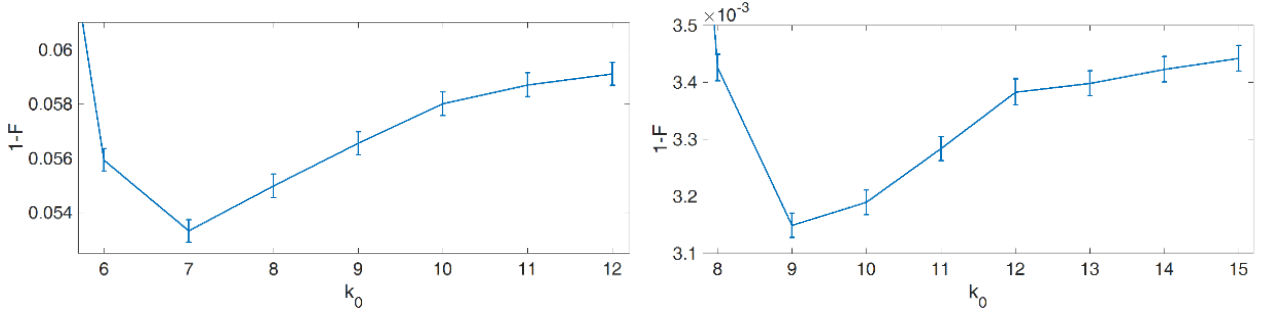

Figure 1. The dependence of fidelity loss from AQFT depth. On the left: number of qubits $n = 12$, the noise level $e = 0.05$. On the right: number of qubits $n = 15$, the noise level $e = 0.01$.

## 5. APPROBATION BASED ON THE PÖSCHL-TELLER POTENTIAL

This section presents the results of numerical experiments aimed to demonstrating the correctness of the above formulas and methods. For this, we simulated the dynamics of a quantum particle in the Pöschl-Teller potential. This potential is given by the following formula [16,17]:

$$V(x) = -\frac{V_0}{\operatorname{ch}^2(x/a)} = -\frac{\hbar^2}{2ma^2}\frac{\lambda(\lambda-1)}{\operatorname{ch}^2(x/a)}. \quad (10)$$

Here, the dimensionless parameter $\lambda$ characterizes the depth of the potential well. The parameters $V_0$ and $a$ are the quantities that characterize the potential well depth and its spatial scale. The considered problem has exact quantum-mechanical solution [16,17]. Let restrict ourselves to considering the discrete spectrum states (i.e., bound states corresponding to negative or zero energy). The corresponding energy levels are

$$E_n = -\frac{\hbar^2}{2ma^2}(\lambda - 1 - n)^2, \quad (11)$$

where $n = 0, 1, ..., n_{max}$. Note that bound states correspond to the condition: $n_{max} \leq \lambda - 1$.

The wave (unnormalized) functions of stationary states, expressed in terms of the hypergeometric function, are

$$\psi_n(x) = \frac{1}{\operatorname{ch}^{\lambda-1-n}(x/a)} F\left[-n, 2\lambda-1-n, \lambda-n, \frac{1}{2}(1-\operatorname{th}(x/a))\right], \quad (12)$$

where $-\infty < x < \infty$. Another notation (in terms of Gegenbauer polynomials, which are a particular form of Jacobi polynomials) has the form

$$\psi_n(x) = \frac{1}{\operatorname{ch}^{\lambda-1-n}(x/a)} C_n^{\lambda-n-\frac{1}{2}}(\operatorname{th}(x/a)). \quad (13)$$

The results of numerical calculations by the Zalka-Wisner method taking into account quantum noise and the exact analytical solution of the nonstationary Schrödinger equation are compared in examples below.

The superposition of the zero and first modes (the potential well depth and the spatial scale were chosen $\lambda = 4$ and $a = 1$ respectively) state was considered:

$$\psi(x,t) = \frac{1}{\sqrt{2}}\left[\exp\left(-i\frac{E_0 t}{\hbar}\right)\psi_0(x) + i\exp\left(-i\frac{E_1 t}{\hbar}\right)\psi_1(x)\right]. \quad (14)$$

Fidelity metric $F = |\langle \psi_{noise} | \psi_{theor} \rangle|^2$ was considered to characterize the accuracy. This metric characterizes the correspondence between the noisy wave function $\psi_{noise}$ obtained by the Zalka-Wiesner method and the exact (but discretized) solution $\psi_{theor}$.

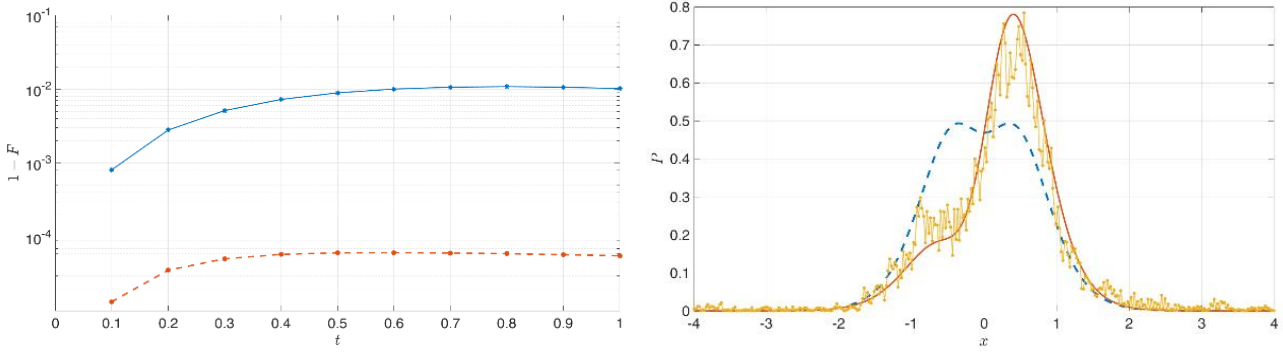

Figure 2. On the left: the dependence of the fidelity loss from the evolution time of the quantum state for the default (upper solid line) and modified Trotter formulas (lower dashed line). On the right: an illustration of the distribution density evolution in the coordinate representation (the initial state is the dashed curve, the final state at the time $t=1$ is the solid curve, the noisy Zalka-Wiesner solution is represented by a set of points).

Figure 2 (left) shows a comparison of the default and modified Trotter formulas for a 7 qubits system and $\Delta t = 0.1$. It can be seen that the considered modified Trotter formula provides an error that is more than 2 orders lower than the error of the standard Trotter formula. Figure 2 (right) gives a visual representation of the quantum noise influence level on the accuracy of obtaining a Schrödinger equation solution on a quantum computer. The evolution of the 9 qubits quantum state was considered over a time interval with a time step $\Delta t = 0.05$ and noise level $e = 0.01$.

A comparison of the developed theory of the quantum noise influence on the simulation accuracy with the results of numerical calculations is shown in Figure 3. It can be seen from this figure (on the left) that the improved estimate based on formula (9) has much better accuracy than the estimation based on (7) and visually coincides with numerical experiments. The evolution of the 7 qubits quantum state was considered on a time interval $0 \leq t \leq 1$ with a time step $\Delta t = 0.05$ and noise level $e = 0.01$. Overall 30 experiments were performed. On the right side of Figure 3, we can see that the improved estimation gives an adequate fidelity prediction for different numbers of qubits in the register. The evolution of the quantum state registers with 7, 8, and 9 qubits was considered over a time interval $0 \leq t \leq 1$ and time step $\Delta t = 0.05$ and noise level $e = 0.01$. Overall 30 experiments were performed for each register.

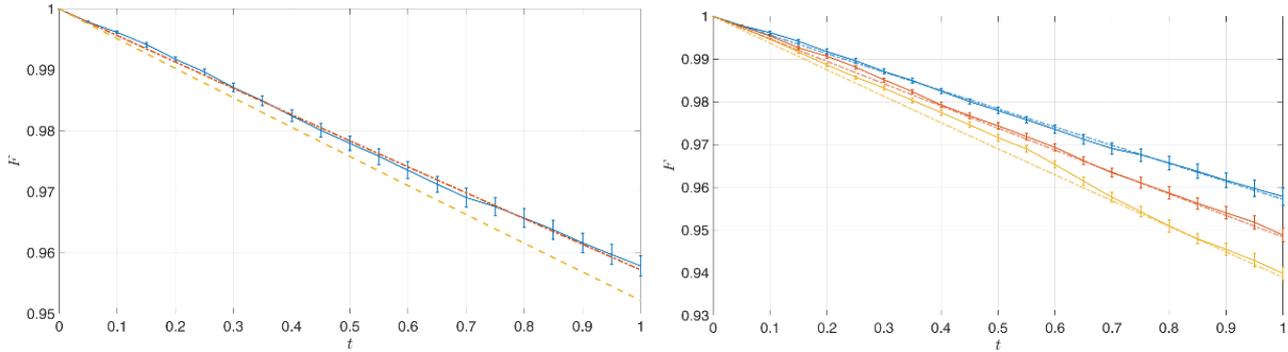

Figure 3. On the left: the dependence of the Zalka-Wiesner algorithm accuracy for modeling the motion of a particle in the Pöschl-Teller potential from the state evolution time (the initial theoretical estimate is the dashed line, the improved estimate is the dash-dotted line, the noisy numerical solution of the Zalka-Wiesner is the solid line with marked rms errors). On the right: the dependence of the Zalka-Wiesner algorithm accuracy for modeling the motion of a particle in the Pöschl-Teller potential from the state evolution time using a different number of qubits in the quantum register (noisy numerical solution of Zalka-Wiesner are solid lines with marked rms errors, improved theoretical estimates are dashed lines, curves from top to bottom correspond to 7, 8 and 9 qubits, respectively).

In the cases shown in Figure 3, the sampling error turned out to be negligible compared to the errors occured by noise in the quantum Fourier transform due to the finite number of qubits in the register. For this reason, the fidelity level $F$ turned out to be lower for a larger number of qubits in the register. The theoretical curves in Figure 3 are plotted according to the following formula based on formula (9):

$$F = \left(F_{QFT}\right)^{2t/\Delta t}. \qquad (15)$$

The value $2t/\Delta t$ represents the number of time steps in formula (15). The coefficient 2 in the exponent is associated with the fact that the Fourier transform is performed twice (direct and inverse transformations) at each step.

Figure 4 shows the prediction of solving the Schrödinger equation accuracy for many-electron systems with different numbers of electrons and different noise levels. The prediction was obtained by formula (15) at $t=1$, $\Delta t = 0.1$. It was assumed that the system of electrons is specified in the configuration space, while $n_0 = 8$ qubits were allocated for each coordinate. The corresponding Fourier transform of each coordinate contains $n_0$ Hadamard transforms and $n_0(n_0-1)/2$ controlled phase shift transforms.

Figure 4 shows that an error level $e=0.01$ or lower is required for adequate modeling of quantum systems of tens of electrons. For modeling systems of hundreds of electrons, it is necessary to provide a level $e=0.001$ or lower. In this case, the error probability of an individual gate should reach the level $e^2$, i.e. $10^{-4}$-$10^{-6}$ order.

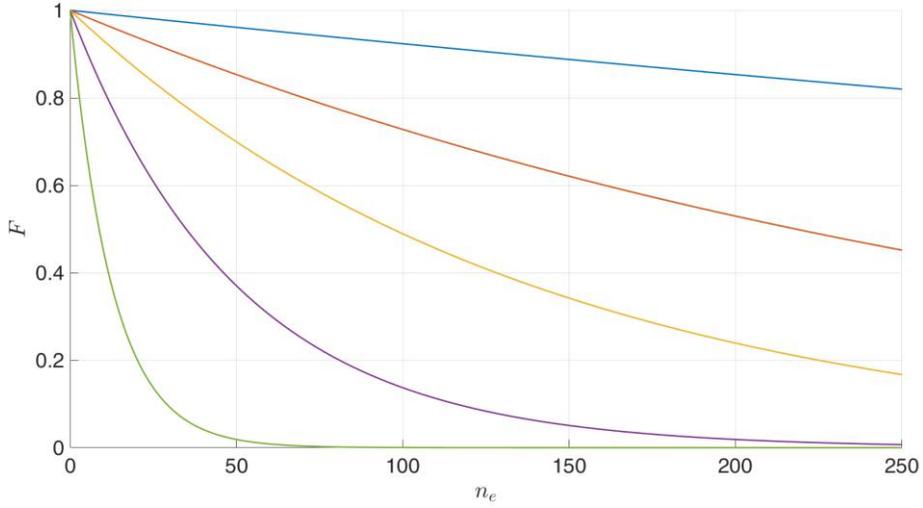

Figure 4. Dependence of the Zalka-Wiesner algorithm simulation accuracy for many-electron systems from the number of electrons and various noise levels. The curves from top to bottom correspond to increasing values of the noise level, respectively $e=0.001$, $e=0.002$, $e=0.003$, $e=0.005$, $e=0.01$.

Note that modeling can be successful even at relatively low levels of accuracy, when $F$ is about 0.01 or even lower. In this case, the desired simulation result should be achieved due to the large amount of statistics obtained from quantum measurements. The characterizing and debugging of experimental and technological procedures requires extensive usage of quantum tomography methods [18].

We also note that the problem of modeling a quantum system can be combined with carrying out quantum measurements in mutually complementary coordinate and momentum spaces in terms of the tomographic Monte Carlo method proposed in [19].

The Zalka-Wiesner algorithm can be implemented on a variety of platforms, including ions in traps [20], atoms in traps [21, 22], and superconducting qubits [23]. The presented results can be used in various experimental implementations of the algorithm with large number of qubits as a basic model for assessing the final accuracy of the quantum algorithm. The multi-qubit implementation of the algorithm makes it possible to find solutions to the Schrödinger equation for very complex quantum systems, radically expanding the capabilities of modern analytical and numerical methods for solving this equation [24].

In this study, we focused on analyzing the noise influence in quantum Fourier transform on the solving the Schrödinger equation accuracy, ignoring several significant factors, such as the finite sampling accuracy, errors in phase transforms, relativistic corrections, the fermionic nature of many-electron systems, etc. Analysis of these factors influence on the accuracy of solving the Schrödinger equation may become the subject of next works.

# 6. CONCLUSIONS

An approach allows us to simulate quantum systems on a quantum computer using the Zalka-Wiesner method taking into account quantum noise has been developed.

Analytical estimates for the accuracy of the Zalka-Wiesner algorithm, due to errors in the implementation of the quantum Fourier transform are obtained.

The developed methods and algorithms are tested on the example of solving the nonstationary Schrödinger equation for a particle in the Pöschl-Teller potential. Good agreement of the results of numerical simulations by the Monte Carlo method with analytical estimates is demonstrated.

A method for predicting the accuracy of solving the Schrödinger equation for many-electron quantum systems is presented. The authors hope that the results of this study are essential for the development of methods for modeling quantum systems using quantum computers and simulators, which is critically important for solving practically significant problems in various scientific fields.

# ACKNOWLEDGMENTS

This work was supported by the Ministry of Science and Higher Education of the Russian Federation (program no. FFNN-2022-0016 for the Valiev Institute of Physics and Technology, Russian Academy of Sciences), and by the Foundation for the Advancement of Theoretical Physics and Mathematics BASIS (project no. 20-1-1-34-1).